# Stability and Cost Optimization in Controlled Random Walks Using Scheduling Fields

Gerhard Wunder[1], Chan Zhou[2], Martin Kasparick[3]

[1]Technische Unversität Berlin, Germany (*gerhard.wunder@hhi.fraunhofer.de*)
[2]Huawei Technologies, Munich, Germany (*chan.zhou@huawei.com*)
[3]Technische Unversität Berlin, Germany (*martin.kasparick@mk.tu-berlin.de*)

**Abstract.** The control of large queueing networks is a notoriously difficult problem. Recently, an interesting new policy design framework for the control problem called h-MaxWeight has been proposed: h-MaxWeight is a natural generalization of the famous MaxWeight policy where instead of the quadratic any other surrogate value function can be applied. Stability of the policy is then achieved through a perturbation technique. However, stability crucially depends on parameter choice which has to be adapted in simulations. In this paper we use a different technique where the required perturbations can be directly implemented in the weight domain, which we call a scheduling field then. Specifically, we derive the theoretical arsenal that guarantees universal stability while still operating 'close' to the underlying cost criterion. Simulation examples suggest that the new approach to policy synthesis can even provide significantly higher gains irrespective of any further assumptions on the network model or parameter choice.

**1. Introduction.** Control policy synthesis for stochastic queueing networks has a multitude of practical applications ranging from the Internet and other data networks over transport networks, manufacturing networks, and power distribution networks in industry to mobile ad-hoc networks. Typically, there are two underlying design criteria: 1) throughput optimality (i.e. the policy supports every set of arrival rates which can potentially be supported by any other algorithm) and 2) optimality with respect to some predefined cost criterion. In the literature there exist a vast number of control policies for queueing networks. One famous approach is the MaxWeight policy, originally introduced by Tassiulas and Ephremides in [9], which is known to be throughput optimal. However, it is seldom used in practice in its pure form since it potentially leads to large delays. Even worse, under low load MaxWeight can behave entirely irrational looping single packets for a long period of time. An example network (originally presented in [8]) where MaxWeight shows such behavior is given in Fig. 1.1. Here, a control policy that is designed to minimize delay is expected not to route packets over the small loop. A lot of work was carried out regarding the issue of delay reduction in backpressure based policies (e.g. [1][11][10]), a general class of throughput optimal policies with improved delay performance is presented recently in [7]. A survey on policy synthesis techniques can be found for example in [4].

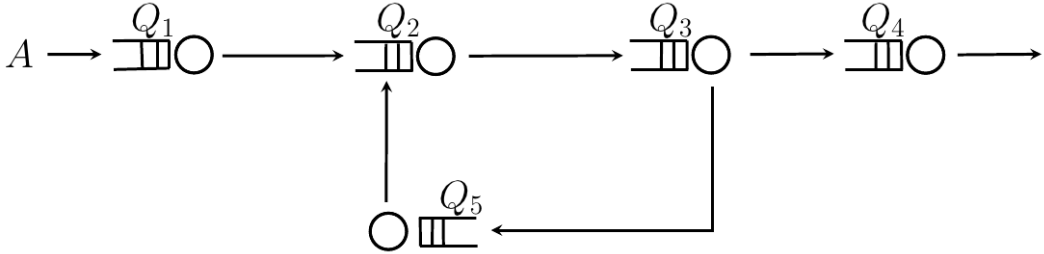

FIG. 1.1. *Exemplary network where MaxWeight shows poor delay performance*

Recently, an interesting new framework for policy design called h-MaxWeight has been proposed in [5][2] which combines the MaxWeight philosophy with a cost criterion. The h-MaxWeight can be seen as a Myopic policy where instead of the quadratic any other cost function can be applied. Stability of the policy is then achieved by perturbing the arguments of the cost function appropriately while still being reasonable 'close' to the original cost function. However, as already mentioned in [5] stability and also cost performance crucially depend on parameter choice which then has to be adapted in simulations. The latter point motivates a different perturbation technique proposed in this paper. It rests upon the observation that apart from technical stability arguments there is actually no specific reason to apply the perturbation technique solely in the argument of the cost function. Instead, when directly applied to the weights the approach becomes much more flexible and, in sharp contrast to [5], can guarantee universal stability while still maintaining asymptotic cost optimality with, in addition, even better cost performance in the non-asymptotic regime. Universal stability and high traffic asymptotic cost optimality analysis becomes involved though since there is in general no longer a natural Ljapunov candidate from the cost function; the theoretical arsenal to circumvent this technical challenge is established in this paper.

*Notation.* We use boldface letters to denote vectors and matrices and common letters with subscript are the elements, such that $A_i$ is the $i$'th element of vector $\boldsymbol{A}$ and $B_{ij}$ is the element in row $i$ and column $j$ of matrix $\boldsymbol{B}$. Moreover $\boldsymbol{A}^T$ refers to the transpose of $\boldsymbol{A}$. $\mathbb{E}\{X\}$ denotes the expected value of random variable $X$. Let $\boldsymbol{I}$ denote the identity matrix of appropriate dimension. Furthermore we denote $\boldsymbol{1}$ the vector of all ones. $\mathbf{diag}(a_1, a_2, ...)$ refers to a diagonal matrix built from the elements $a_1, a_2, ...$ and $\|\cdot\|_i$ denotes the $l_i$ vector norm and $\|\mathbf{x}\|$ is an arbitrary norm. Furthermore we use $\mathcal{A}^c$ to denote the complement of a set $\mathcal{A}$. The probability of $\mathcal{A}$ is denoted as $\Pr\{\mathcal{A}\}$. The indicator $\mathbb{I}\{\cdot\}$ equals 1 if the argument is true and equals 0 otherwise.

**2. System Model.** Similar to [5] we use a simple stochastic network model: the Controlled Random Walk (CRW) model. We consider a queueing network with $m$ queues in total representing $m$ physical buffers with unlimited storage capacity. We arrange the queue backlog in the vector $\mathbf{Q}$, such that $\mathbf{Q} = [Q_1, \ldots, Q_m]^T$ which we refer to as the queue state. Let $\mathcal{M}$ be the set of queue indices. Suppose that the evolution of the queueing system is time slotted with $t \in \mathbb{N}_0$. Then, the CRW model is defined by queueing law:

$$\mathbf{Q}(t+1) = [\mathbf{Q}(t) + \mathbf{B}(t+1)\mathbf{U}(t)]^+ + \mathbf{A}(t+1) \tag{2.1}$$

Here, the vector process $\mathbf{A}(t) \in \mathbb{N}_0^m$ is the (exogenous) influx to the queueing system with mean $\boldsymbol{\alpha} \in \mathbb{R}_+^m$ (vector of arrival rates in packets per slot); $\mathbf{B}(t) \in \mathbb{Z}_0^{m \times l}$ is a matrix process with average $\mathbf{B} \in \mathbb{Z}_0^{m \times l}$, containing both information about network topology (that is, connectivity or routing paths) and service rates[1]. The control $\mathbf{u} = \mathbf{U}(t)$ in slot $t$ is an element of the set $\{0,1\}^l$ constrained by $\boldsymbol{Cu} \leq \mathbf{1}$ using the binary constituency matrix $\mathbf{C} \in \mathbb{Z}_0^{l_m \times l}$ (with $l_m > 0$ being the number of resource constraints in the network). For the sake of notational simplicity we omit the time index in the following where possible. Throughout the entire paper $\boldsymbol{x} \in \mathbb{N}_0^m$ denotes the actual backlog.

In what follows, the queueing system (2.1) is assumed to be a $\delta_{\mathbf{0}}$-irreducible Markov chain ($\delta_{\mathbf{0}}$ being the point measure at $\boldsymbol{x} = \mathbf{0}$).

**2.1. Example.** Consider the introductory example of Fig. 1.1. We have traffic arriving at queue $Q_1$ and leaving the network after being processed at $Q_4$. Moreover in each time slot traffic from queue $Q_3$ can be routed either to $Q_4$ or to $Q_5$, not both, thus $u_3 + u_4 \leq 1$. We assume arrivals with rate $\alpha > 0$ and set all service rates equal to 1 (thus for the network to be stabilizable it is required that $\alpha < 1$). The corresponding CRW model is given by:

$$\mathbf{B} = \begin{bmatrix} -1 & 0 & 0 & 0 & 0 & 0 \\ 1 & -1 & 0 & 0 & 1 & 0 \\ 0 & 1 & -1 & -1 & 0 & 0 \\ 0 & 0 & 1 & 0 & 0 & -1 \\ 0 & 0 & 0 & 1 & -1 & 0 \end{bmatrix}$$

and

$$\mathbf{C} = \begin{bmatrix} 1 & 0 & 0 & 0 & 0 & 0 \\ 0 & 1 & 0 & 0 & 0 & 0 \\ 0 & 0 & 1 & 1 & 0 & 0 \\ 0 & 0 & 0 & 0 & 0 & 1 \\ 0 & 0 & 0 & 0 & 1 & 0 \end{bmatrix}.$$

Let us present some preliminary results next.

**3. Preliminaries.**

**3.1. Stability.** The stability of an $\delta_{\mathbf{0}}$-irreducible Markov chain can be defined in different manners. We first recall the definition of *recurrent* Markov chain as given in [6] based on the measure of the occupation time

$$\eta_{\mathcal{A}} := \sum_{t=1}^{\infty} \mathbb{I}(\mathbf{Q}(t) \in \mathcal{A})$$

which gives the number of visits in a set $\mathcal{A} \subset \mathbb{R}_+^m$ by a Markov chain after time zero. A Markov chain is *recurrent*, if it holds $\mathbb{E}\{\eta_{\mathcal{A}}\} = +\infty$, for any set $\mathcal{A} \subset \mathbb{R}_+^m$. Additionally, if the Markov chain admits an invariant probability measure, then it is *positive recurrent*. If the $\delta_0$-irreducible Markov chain is positive recurrent, it is also *weakly stable* [3] so that it holds

$$\lim_{t \to +\infty} \Pr(\|\mathbf{Q}(t)\| > B(\epsilon)) < \epsilon$$

for any $\epsilon > 0$ and some constant $B(\epsilon) > 0$. In this paper we also apply the following stability definition.

---

[1]If not stated otherwise, service rates are usually assumed to be one throughout the paper.

DEFINITION 3.1. *A Markov chain is called* f-stable, *if there is an unbounded function $f : \mathbb{R}_+^m \to \mathbb{R}_+$, such that for any $0 < B < +\infty$ the set $\mathcal{B} := \{\mathbf{x} : f(\mathbf{x}) \leq B\}$ is compact, and furthermore it holds*

$$\limsup_{t \to +\infty} \mathbb{E}\{f(\mathbf{Q}(t))\} < +\infty. \tag{3.1}$$

In the definition the function $f$ is unbounded in all positive directions so that $f(\mathbf{x}) \to \infty$ if $\|\mathbf{x}\| \to \infty$. Choosing directly $f(\mathbf{x}) = \|\mathbf{x}\|$, Definition 3.1 is equivalent to the definition of *strongly stable* [3] which implies weak stability. Clearly, for any $f(\mathbf{x})$ which grows faster than $\|\mathbf{x}\|$, inequality (3.1) implies that the Markov chain is strongly stable. We call a vector of arrival rates $\boldsymbol{\alpha} \in \mathbb{R}_+^m$ *stabilizable* when the corresponding queueing system driven by some specific scheduling policy is positive recurrent.

A scheduling policy is now called *throughput optimal* if it keeps the Markov chain positive recurrent for any vector of arrival rates $\boldsymbol{\alpha}$ for which a stabilizing policy exists. It is easy to show that for our model, by introducing the velocity set

$$\mathcal{V} := \{\mathbf{v} \in \mathbb{R}_+^m, \mathbf{v} = \mathbf{B}\mathbf{u} + \boldsymbol{\alpha}\},$$

the system is stabilizable if and only if the interior of $\mathcal{V}$ contains $\mathbf{v} = \mathbf{0}$ [5].

**3.2. Myopic Policies: h-MaxWeight.** Let us introduce a cost function

$$c : \mathbb{N}_0^m \to \mathbb{R}_+, \mathbf{x} \hookrightarrow c(\mathbf{x}),$$

assigning any queue state a non-negative number. Typically, the goal is to minimize the average cost over a given finite or infinite time period or some discounted cost criterion. The optimal solution to the resulting problems –which in discrete time can be modelled as a *Markov Decision Problem*– can be found by dynamic programming, which is, however, infeasible for large networks.

A simple approach to queueing network control is the *myopic or greedy policy*. Such a policy selects the control decision that minimizes the expected cost only for the next time slot. Many policies can be considered myopic: for example it is shown in [9] that taking $c(\mathbf{x}) = \mathbf{x}^T \mathbf{D} \mathbf{x}$, for some positive diagonal matrix $\mathbf{D}$, the corresponding MaxWeight policy is throughput optimal. However, very little is known about stability properties of other cost function families.

In [5], a cost function based policy design framework called $h$-MaxWeight is introduced which is a generalization of the MaxWeight policy. Meyn considers a slightly different definition of the CRW model, which is characterized by queueing law:

$$\mathbf{Q}(t+1) = \mathbf{Q}(t) + \mathbf{B}(t+1)\mathbf{U}(t) + \mathbf{A}(t+1) \tag{3.2}$$

The control $\mathbf{U}(t) \in \mathbb{N}_0^l$ is an element of the region

$$\mathcal{U}^*(\mathbf{x}) := \mathcal{U}(\mathbf{x}) \cap \{0,1\}^l,$$

with

$$\mathcal{U}(\mathbf{x}) := \left\{\mathbf{u} \in \mathbb{R}_+^l : \mathbf{C}\mathbf{u} \leq \mathbf{1}, [\mathbf{B}\mathbf{u} + \boldsymbol{\alpha}]_i \geq 0 \text{ for } x_i = 0\right\}.$$

In the $h$-MaxWeight based control policy, the control vector is derived according to

$$\underset{\boldsymbol{u} \in \mathcal{U}^*(\mathbf{x})}{\arg\min} < \nabla h(\boldsymbol{x}), \boldsymbol{B}\boldsymbol{u} + \boldsymbol{\alpha} > . \tag{3.3}$$

Thus, the policy is myopic with respect to the gradient of some perturbation $h$ of the underlying cost function. Meyn develops two main constraints on the function $h$: the first requires the partial derivative of $h$ to vanish when queues become empty:

$$\frac{\partial h}{\partial x_i}(\boldsymbol{x}) = 0 \quad \text{if } x_i = 0 \tag{3.4}$$

Moreover the dynamic programming inequality has to hold for the function $h$:

$$\min_{\boldsymbol{u} \in \mathcal{U}(\boldsymbol{x})} <\nabla h(\boldsymbol{x}), \boldsymbol{Bu} + \boldsymbol{\alpha}> \leq -c(\boldsymbol{x}) \tag{3.5}$$

When $h$ is non-quadratic, the derivative condition (3.4) is not always fulfilled. Therefore a perturbation technique is used where $h(\boldsymbol{x}) = h_0(\tilde{\boldsymbol{x}})$, hence it is a perturbation of a function $h_0$. Two perturbations are proposed: an exponential perturbation with $\theta \geq 1$ given by

$$\tilde{x}_i := x_i + \theta \left(e^{-\frac{x_i}{\theta}} - 1\right), \tag{3.6}$$

and a logarithmic perturbation with $\theta > 0$ defined as

$$\tilde{x}_i := x_i \log\left(1 + \frac{x_i}{\theta}\right). \tag{3.7}$$

While the first approach shows better performance in simulations, the stability of the resulting policy depends on the parameter $\theta$ being sufficiently large (determined by the considered network setting). This is overcome by the second perturbation which is stabilizing for each feasible $\theta$, however it comes with the additional constraint

$$\frac{\partial h_0}{\partial x_i}(\boldsymbol{x}) \geq \epsilon x_i, \quad \forall i \in \mathcal{M}, \tag{3.8}$$

which is a significant limitation on the space of functions that can be chosen as $h_0$. Regarding the stability of the $h$-MaxWeight policy, Meyn devised the following theorem.

THEOREM 3.2 (Theorem 1.1 from [5]). *Consider the model (3.2) satisfying the following conditions:*
  1. *The i.i.d. process $(\boldsymbol{A}, \boldsymbol{B})$ has integer entries, and a finite second moment.*
  2. *$B_{ij}(t) \geq -1$ for each $i,j$ and $t$, and for each $j \in \{1, \ldots, l_u\}$ there exists a unique value $i_j \in \{1, \ldots, l\}$ satisfying*

$$B_{ij}(t) \geq 0 \quad a.s. \ \forall i \neq i_j.$$

  3. *The function $h_0 : \mathbb{R}^m \to \mathbb{R}_+$ satisfies the following:*
     *(a) Smoothness: The gradient $\nabla h_0$ is Lipschitz continuous,*
     *(b) Monotonicity: $\nabla h_0(\boldsymbol{x}) \in \mathbb{R}_+^m$ for $\boldsymbol{x} \in \mathbb{R}_+^l$,*
     *(c) The dynamic programing inequality (3.5) holds, with $c$ a norm on $\mathbb{R}^l$.*

*Then, there exists $\theta_0 < \infty$ and $\bar{\eta}_h < \infty$ such that for any $\theta \geq \theta_0$, the following bound holds under the $h$-MaxWeight policy with perturbation (3.6):*

$$\mathbb{E}\left\{h(\boldsymbol{Q}(t+1)) - h(\boldsymbol{Q}(t))|\boldsymbol{Q}(t) = \boldsymbol{x}\right\} \leq -\frac{1}{2}c(\boldsymbol{x}) + \frac{1}{2}\bar{\eta}_h$$

*Consequently, it is:*

$$n^{-1}\mathbb{E}\left\{\sum_{t=0}^{n-1} c(\boldsymbol{Q}(t))\right\} \leq 2n^{-1}h(\boldsymbol{x}) + \bar{\eta}_h, \quad n \geq 1, \boldsymbol{x} \in \mathbb{Z}_+^l$$

Already in [5], Meyn mentioned some of the drawbacks of this $h$-MaxWeight policy: it depends crucially on parameter choice and is therefore not throughput optimal (which actually motivated the second perturbation (3.7)). Furthermore, the approach also depends on additional constraints on the network topology (cf. Theorem 3.2, Condition 2)) so that it is not easily applicable to more general networks. A different approach is described next: note at first that it is by no means necessary to choose the perturbation as in (3.6) as long as we stay reasonably 'close' to $h_0$ while still maintaining stability (the only argument is technical because there is typically no longer a natural Ljapunov candidate obtained from $h_0$). By contrast here, we directly start with the function $\boldsymbol{\mu}(\boldsymbol{x}) := \nabla h(\boldsymbol{x})$ and derive conditions properties that guarantee throughput optimality, irrespective of the actual parameters chosen.

### 4. $\mu$-MaxWeight Network Control.

**4.1. Sufficient Stability Conditions.** In this section, we give generalized sufficient conditions for throughput optimality for the systems (2.1), (3.2). In what follows, we consider scheduling policies of the form

$$\mathbf{u}^*(\mathbf{x}) = \operatorname*{arg\,min}_{\mathbf{u} \in \mathbb{R}^n_+ : C\boldsymbol{u} \leq \mathbf{1}} \langle \boldsymbol{\mu}(\mathbf{x}), \mathbf{B}\mathbf{u} + \boldsymbol{\alpha} \rangle, \tag{4.1}$$

where $\boldsymbol{\mu}(\mathbf{x})$ is a vector valued function $\mathbb{R}^m_+ \to \mathbb{R}^m_+$, which is called the *weight vector* for some actual queue state $\mathbf{x}$. Note that $\boldsymbol{\mu}$ is reminiscient of a vector field and can thus be interpreted as a *scheduling field* for which we present a stability characterization. Observe that by construction of the policy we can normalize the weight vector as

$$\bar{\boldsymbol{\mu}}(\mathbf{x}) := \frac{\boldsymbol{\mu}(\mathbf{x})}{\|\boldsymbol{\mu}(\mathbf{x})\|_1} \tag{4.2}$$

without loss of generality and hence $\|\bar{\boldsymbol{\mu}}(\mathbf{x})\|_1 = 1$. Furthermore, we assume that the resulting policy is non-idling, i.e. $\|\boldsymbol{\mu}(\mathbf{x})\|_1 = 0$ if and only if $\mathbf{x} = \mathbf{0}$.

THEOREM 4.1. *Consider the queueing system (2.1) driven by the control policy (4.1) with some scheduling field $\boldsymbol{\mu}$. The policy is throughput optimal if the corresponding normalized scheduling field given in Eqn. (4.2) fulfills the following conditions:*
  1. *Given any $0 < \epsilon_1 < 1$ and $C_1 > 0$, there is some $B_1 > 0$ so that for any $\Delta\mathbf{x} \in \mathbb{R}^m$ with $\|\Delta\mathbf{x}\| < C_1$, we have $|\bar{\mu}_i(\mathbf{x} + \Delta\mathbf{x}) - \bar{\mu}_i(\mathbf{x})| \leq \epsilon_1$ for any $\mathbf{x} \in \mathbb{R}^m_+$ with $\|\mathbf{x}\| > B_1$, $\forall i \in \mathcal{M}$.*
  2. *Given any $0 < \epsilon_2 < 1$ and $C_2 > 0$, there is some $B_2 > 0$ so that for any $\mathbf{x} \in \mathbb{R}^m_+$ with $\|\mathbf{x}\| > B_2$ and $x_i < C_2$, we have $\bar{\mu}_i(\mathbf{x}) \leq \epsilon_2$, for any $i \in \mathcal{M}$.*

*Moreover, for any stabilizable arrival process the queueing system is f-stable under the given policy where f is an unbounded function as defined in Definition 3.1. The exact formulation of f depends on the field $\bar{\boldsymbol{\mu}}(\mathbf{x})$.*

*Proof.* The proof is given in Appendix A. □

Remarkably, there is no a priori need for the dynamic programming inequality (it follows). Theorem 4.1 can be further refined and tailored to the situation in Theorem 3.2.

COROLLARY 4.2. *Consider the queueing system (3.2) driven by the control policy (3.3) with some cost function h. Suppose the corresponding scheduling field $\boldsymbol{\mu}(\boldsymbol{x}) := \nabla h(\boldsymbol{x})$ is continuously differentiable and Condition 2) in Theorem 3.2 on network topology $\{\boldsymbol{B}(\cdot)\}$ holds. Then, the following conditions are sufficient for throughput optimality:*
  1. *For any $\epsilon > 0$ there is some $C_1^* > 0$ so that for all $\|\boldsymbol{x}\| \geq C_1^*$:*

$$\|\nabla \log(\mu_i(\boldsymbol{x}))\| \leq \epsilon, \quad \forall i \in \mathcal{M}$$

  2. *If $x_i = 0$ then $\mu_i(\boldsymbol{x}) = 0$, $\forall i \in \mathcal{M}$.*

*Proof.* The proof can be found in Appendix B. □

REMARK 1. *The restriction on the network topology in Corollary 4.2 can be omitted if Condition 2) is replaced with Condition 2) of Theorem 4.1.*

Theorem 4.1 can also be tailored to policies with simple perturbations (by simple we mean each component is perturbed by a single parameter).

COROLLARY 4.3. *Suppose, everything is as in Corollary 1. Let the scheduling field be defined as $\boldsymbol{\mu}(\boldsymbol{x}) := \nabla h_0(\tilde{\boldsymbol{x}})$ for some given simple perturbation $\tilde{\boldsymbol{x}}$. Then, for some $\epsilon > 0$,*

$$\frac{\partial \tilde{x}_i}{\partial x_i} \text{ is Lipschitz, and } \frac{\partial \tilde{x}_i}{\partial x_i} \to \infty, x_i \to \infty,$$

$$\frac{\partial h_0}{\partial x_i} \text{ is Lipschitz, and } \frac{\partial h_0}{\partial \tilde{x}_i}(\tilde{\boldsymbol{x}}) \geq \left(\frac{\partial \tilde{x}_i}{\partial x_i}\right)^{1+\epsilon}, x_i \to \infty,$$

*is sufficient for stability.*

*Proof.* The proof can be found in Appendix C. □

REMARK 2. *The conditions in Corollary 4.3 cover indeed a larger class of throughput optimal policies compared e.g. to the perturbation in (3.7) together with condition (3.8) since it is only required that $\frac{\partial h_0}{\partial x_i}$ grows as $\log^{1+\epsilon} x_i$ in each component (observe that $\frac{\partial \tilde{x}_i}{\partial x_i} = \log\left(1 + \frac{x_i}{\theta}\right) + \frac{x_i}{\theta + x_i}$ which is also Lipschitz).*

REMARK 3. *(On extensions) A weaker condition than Condition 1) in Corollary 4.2 is as follows: for any $\epsilon > 0$ there is some $C_1^{**} > 0$ so that for all $\|\boldsymbol{x}\| \geq C_1^{**}$:*

$$\|\nabla \mu_i(\boldsymbol{x})\| \leq \epsilon \|\boldsymbol{\mu}(\boldsymbol{x})\|, \quad \forall i \in \mathcal{M}$$

*We conjecture that this will establish the most general condition. Furthermore, we showed in the wireless broadcast setting that the conditions are also necessary if the boundary of the feasible (rate) region contains differentiable parts, i.e. parts where the hyperplanes defined through the scheduling field are uniquely supported [12].*

**5. Policy Design.** Corollary 4.2 makes it apparent that one is not confined to the specific policy design rule in Theorem 3.2 and can ensure much easier throughput optimality by directly modifying the scheduling field appropriately. Let us demonstrate this by a simple example:

We consider a simple network of two queues in tandem. Assume we have a linear cost function $c(\boldsymbol{x}) = c_1 x_1 + c_2 x_2$. Similar to [5] we perturb the optimal value function from the fluid model $J^*$ (which is known in this setting); thus (assuming parameters as in [5]) we have

$$h_0(\boldsymbol{x}) = J^*(\boldsymbol{x}) = \frac{1}{2} d_1 (x_1 + x_2)^2 + \frac{1}{2} d_2 x_2^2$$

with $d_1 = \frac{c_1}{\nu_2 - \alpha_1}$ and $d_2 \frac{c_2 - c_1}{\nu_2}$. The gradient $\nabla h$ is then given by:

$$\nabla h(\boldsymbol{x}) = \begin{pmatrix} d_1(\tilde{x}_1 + \tilde{x}_2)(1 - e^{-\frac{x_1}{\theta}}) \\ (d_1(\tilde{x}_1 + \tilde{x}_2) + d_2 \tilde{x}_2)(1 - e^{-\frac{x_2}{\theta}}) \end{pmatrix}$$

Thereby, the exponential perturbation (3.6) was used. The $h$-MaxWeight policy which we obtain using this function in (3.3) does not fulfill the conditions of Corollary 4.2 for throughput optimality. Therefore we can (intuitively) derive a different perturbation:

$$\boldsymbol{\mu}(\boldsymbol{x}) = \begin{pmatrix} d_1(\tilde{x}_1 + \tilde{x}_2)(1 - e^{-\frac{x_1}{\theta(1+x_2)}}) \\ (d_1(\tilde{x}_1 + \tilde{x}_2) + d_2 \tilde{x}_2)(1 - e^{-\frac{x_2}{\theta(1+x_1)}}) \end{pmatrix}$$

Setting

$$\mathbf{P}_\theta(\mathbf{x}) := \mathbf{diag}\left(1 - \exp\left(-\frac{x_i}{\theta(1 + \sum_{j \neq i} x_j)}\right)\right) \tag{5.1}$$

the policy can be concisely written as $\boldsymbol{\mu}(\boldsymbol{x}) = \mathbf{P}_\theta(\mathbf{x}) \nabla h_0(\tilde{\boldsymbol{x}})$. It can be easily verified that the conditions of Corollary 4.2 hold for suitable $h_0$, i.e. it is throughput optimal for any $\theta > 0$. Observe also that we have incorporated the useful property that queues will not be served when other queues tend to infinity.

**5.1. Numerical Results.** Subsequently, we numerically compare the policies obtained with the $\mu$-MaxWeight framework to MaxWeight and the generalized $h$-MaxWeight. For this we consider the introductory example of Fig. 1.1, described in detail in the context of the CRW model in Section 2.1. As mentioned before, MaxWeight can show bad performance w.r.t. delay especially under low load. We want to evaluate whether we can improve that performance using the cost-function based approach. Since the optimal value function from the fluid model is not readily available we assume the linear cost-function $c(\boldsymbol{x}) = \mathbf{c}^T\mathbf{x}$ for simplicity; hence, the resulting weight vector is $\boldsymbol{\mu}(\boldsymbol{x}) = \mathbf{P}_1(\mathbf{x})\,\boldsymbol{c}$ with $\mathbf{P}_1(\mathbf{x})$ given in (5.1) with $\theta = 1$.

Fig. 5.1(a) and Fig. 5.1(b) depict the average cost for different arrival rates after a running time of 10000 time slots. Fig. 5.1(a) shows the case when all $c_i = 1$. It can be observed that our cost-function based approach provides significant gains over MaxWeight at all arrival rates. As expected the gain increases with decreasing network load. Moreover, we observe a similar performance as the generalized $h$-MaxWeight policies ($\theta$ chosen sufficiently large) even though throughput-optimality is imposed.

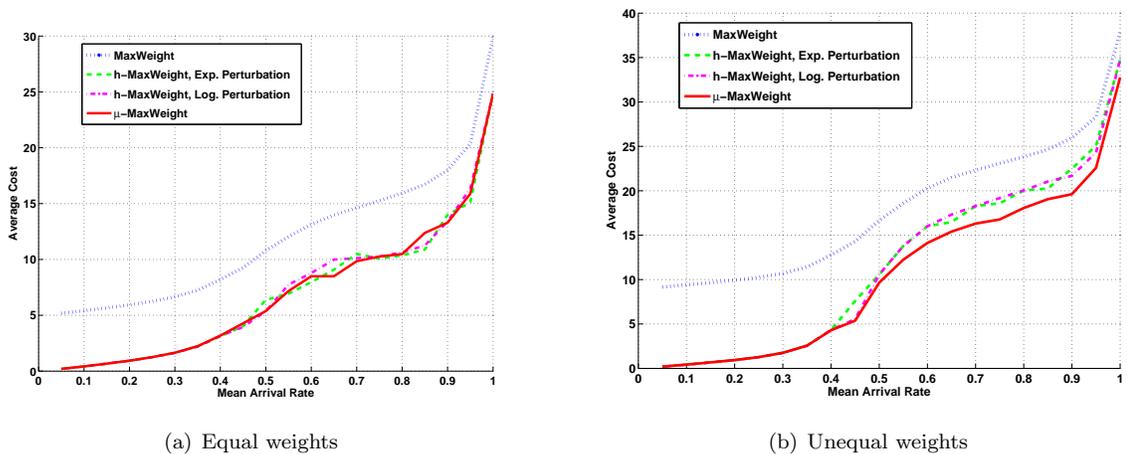

(a) Equal weights  (b) Unequal weights

FIG. 5.1. *Numerical comparison of the average cost of different policies.*

Since our approach is based on a cost metric it is natural to ask how it behaves in case the queues are not weighted equally. For example assume we want to discourage the use of the reverse loop, thus we set $c_5 > c_i$ for all $i \neq 5$. Fig. 5.1(b) compares the control policies assuming $c_5 = 5$. In this case, when the load increases our approach even outperforms $h$-MaxWeight with both exponential and logarithmic perturbation, while at the same time providing throughput optimality.

**6. Conclusion.** We introduced a control policy synthesis framework for queueing networks that combines throughput optimality per design with optimization with respect to an arbitrary cost metric. To design such a policy we derive fundamental theoretical conditions that guarantee universal stability and can easily be checked. We have shown that we can achieve higher performance gains both over classical MaxWeight routing as well as generalized MaxWeight algorithms, however, without the inherent limitations such as parameter dependent stability or additional constraints on the network model.

**Appendix A. Proof of Theorem 4.1.**
Stability can be proven by establishing the so-called *Lyapunov drift criteria* as given in [9, 3]. That is to say if we can find some non-negative $V(\mathbf{x}) : \mathbb{R}_+^m \to \mathbb{R}_+$, some $\theta > 0$ and a compact region $\mathcal{B} := \{\mathbf{x} : \|\mathbf{x}\| \leq B\}$ such that

$$\mathbb{E}\{V(\mathbf{Q}(t+1))|\mathbf{Q}(t)\} < +\infty \qquad \forall \mathbf{Q}(t) \in \mathcal{B} \qquad (A.1)$$
$$\Delta V(\mathbf{Q}(t)) < -\theta \qquad \forall \mathbf{Q}(t) \in \mathcal{B}^c, \qquad (A.2)$$

the queueing system is positive recurrent. Here, $\Delta V\left(\mathbf{Q}\left(t\right)\right)$ is the one step drift defined as

$$\Delta V\left(\mathbf{Q}\left(t\right)\right) := \mathbb{E}\left\{V\left(\mathbf{Q}\left(t+1\right)\right) - V\left(\mathbf{Q}\left(t\right)\right) \middle| \mathbf{Q}\left(t\right)\right\}.$$

Furthermore, if for some $\theta > 0$, it satisfies

$$\Delta V\left(\mathbf{x}\right) \leq -\theta f\left(\mathbf{x}\right), \qquad \forall \mathbf{Q}\left(t\right) \in \mathcal{B}^c, \tag{A.3}$$

for some $B > 0$ and unbounded positive function $f\left(\mathbf{x}\right)$, it can be shown that the queueing system is f-stable.

We carry out the proof in two steps. First, we prove throughput optimality for those policies, whose corresponding fields $\bar{\boldsymbol{\mu}}(\mathbf{x})$ fulfill the integrability condition in Eqn. (A.4) below. The fields in those policies can be regarded as the normalized gradient of a certain potential field $V(\mathbf{x})$. We show that the expected drift $\Delta V\left(\mathbf{x}\right)$ satisfies the inequality (A.3) and hence the system driven by those policies is stable. In the second step, we extend the results to all other policies whose corresponding fields are not integrable. It is shown that if the policies fulfill the condition given in the theorem, their fields $\bar{\boldsymbol{\mu}}(\mathbf{x})$ can be approximated by some $\tilde{\boldsymbol{\mu}}(\mathbf{x})$ which is integrable. Then we prove the drift condition $\Delta V\left(\mathbf{x}\right)$ for those policies and establish the stability.

First, we analyze the subclass of fields whose $\bar{\mu}_i(\mathbf{x})$ are continuously differentiable. Furthermore, we assume that the field satisfy the integrability condition, i.e.,

$$\frac{\partial\left(\bar{\mu}_i\left(\mathbf{x}\right)\right)}{\partial x_j} = \frac{\partial\left(\bar{\mu}_j\left(\mathbf{x}\right)\right)}{\partial x_i}, \qquad \forall i, j \in \mathcal{M}. \tag{A.4}$$

For such scheduling policies, we have the following lemma.

LEMMA A.1. *If Eqn. (A.4) holds for all $\mathbf{x} \in \mathbb{R}_+^m$, then any stabilizable vector of arrival rates $\boldsymbol{\alpha}$ is also stabilizable under the corresponding scheduling policy as long as $\bar{\boldsymbol{\mu}}(\mathbf{x})$ fulfills the conditions given in Theorem 4.1.*

*Proof.* Condition (A.4) implies that the vector field defined by $\bar{\boldsymbol{\mu}}(\mathbf{x})$ has the path independence property, namely the integral of $\bar{\boldsymbol{\mu}}(\mathbf{x})$ along a path depends only on the start and end points of that path, not the particular route taken. According to *Poincaré lemma*, the vector field $\bar{\boldsymbol{\mu}}(\mathbf{x})$ is completely integrable and it is the gradient of a scalar field, that is to say, there exist some function $f(\mathbf{x}) : \mathbb{R}_+^m \to \mathbb{R}_+$ with

$$\frac{\partial f(\mathbf{x})}{\partial x_i} = \bar{\mu}_i(\mathbf{x}). \tag{A.5}$$

Setting the value of $f(\mathbf{x})$ at the origin equal zero, $f(\mathbf{x})$ at the point $\mathbf{x}$ can be calculated by

$$f(\mathbf{x}) = \int_0^{\|\mathbf{x}\|_2} \bar{\boldsymbol{\mu}}\left(t\bar{\mathbf{x}}\right)^T \bar{\mathbf{x}} dt, \tag{A.6}$$

where $\bar{\mathbf{x}} := \frac{\mathbf{x}}{\|\mathbf{x}\|_2}$ is the normalized vector of $\mathbf{x}$. Since each element of $\bar{\boldsymbol{\mu}}(\mathbf{x})$ is larger than or equal to zero, $f(\mathbf{x})$ is a positive function. Moreover, if $\|\mathbf{x}\|$ becomes large, according to Condition 2) in the Theorem 4.1, for $i$-th queue with bounded $x_i$, $\bar{x}_i \to 0$ results in $\bar{\mu}_i(\mathbf{x}) \to 0$. Then for other queues with $\bar{\mu}_j(\mathbf{x}) > C_\mu$, $x_j$ grows with increasing $\|\mathbf{x}\|$ and we have $\bar{x}_j > C_x$ for some $C_x$ and $C_\mu > 0$. Thus it holds

$$\bar{\boldsymbol{\mu}}(\mathbf{x})^T \bar{\mathbf{x}} > C$$

for some $C > 0$ if $\|\mathbf{x}\|$ is sufficiently large. Considering Eqn. (A.6), it follows that $f(\mathbf{x}) \to +\infty$ as $\|\mathbf{x}\| \to +\infty$. Therefore, $f(\mathbf{x})$ is a positive, unbounded function as we used in Definition 3.1.

Observing a new vector field defined by $\boldsymbol{\nu}(\mathbf{x}) = f(\mathbf{x})\bar{\boldsymbol{\mu}}(\mathbf{x})$, we have

$$\begin{aligned}
\frac{\partial(\nu_i(\mathbf{x}))}{\partial x_j} &= \frac{\partial(f(\mathbf{x})\bar{\mu}_i(\mathbf{x}))}{\partial x_j} \\
&= \bar{\mu}_j(\mathbf{x})\bar{\mu}_i(\mathbf{x}) + \frac{\partial \bar{\mu}_j(\mathbf{x})}{\partial x_i} \\
&= \bar{\mu}_j(\mathbf{x})\bar{\mu}_i(\mathbf{x}) + \frac{\partial \bar{\mu}_i(\mathbf{x})}{\partial x_j} \\
&= \frac{\partial(f(\mathbf{x})\bar{\mu}_j(\mathbf{x}))}{\partial x_i} \\
&= \frac{\partial(\nu_j(\mathbf{x}))}{\partial x_i}, \qquad \forall i,j \in \mathcal{M}.
\end{aligned} \tag{A.7}$$

Condition (A.7) ensures that $\boldsymbol{\nu}(\mathbf{x})$ is also the gradient of a scalar field and there is a function $V(\mathbf{x}) : \mathbb{R}^m_+ \to \mathbb{R}_+$ with

$$\frac{\partial V(\mathbf{x})}{\partial x_i} = f(\mathbf{x})\bar{\mu}_i(\mathbf{x}),$$

where $f(\mathbf{x})$ is the magnitude of the gradient and $\bar{\boldsymbol{\mu}}(\mathbf{x})$ is the direction of the gradient. Set $V(\mathbf{0}) = 0$ and $V(\mathbf{x})$ at the point $\mathbf{x}$ is

$$V(\mathbf{x}) = \int_0^{\|\mathbf{x}\|_2} f(t\bar{\mathbf{x}})\bar{\boldsymbol{\mu}}(t\bar{\mathbf{x}})^T \bar{\mathbf{x}} dt.$$

It is easy to show that the function $V(\mathbf{x})$ is also a positive, unbounded function. We use the function $V(\mathbf{x})$ as our Lyapunov function in the proof.

Subsequently, let $\boldsymbol{r} \in \mathbb{R}^m_+$ be the vector of network induced arrivals plus departures and $\boldsymbol{a} \in \mathbb{R}^m_+$ be the vector of exogenous arrivals. Moreover, let $\boldsymbol{z} \in \mathbb{R}^m_-$ be the vector of number of excess packets compensating for the case when more packets are attempted to be removed than the queue contains. The first condition of the Lyapunov function given in (A.1) is satisfied as long as $a_i, r_i, \forall i \in \mathcal{M}$, are bounded. Next we analyze the second condition, namely the drift of $V(\mathbf{x})$ of the queueing system.

Using the *mean value theorem* of differential calculus we have for some $\tilde{\mathbf{x}}$ between $\mathbf{x}$ and $\mathbf{x} + \Delta\boldsymbol{x}$ i.e. $\tilde{x}_i = \kappa_i x_i + (1-\kappa_i)x_i + \Delta x_i, \forall i \in \mathcal{M}$, for some $\kappa_i \in [0,1]$:

$$\Delta V(\mathbf{x}) = \mathbb{E}\left\{\sum_{i=1}^m f(\tilde{\mathbf{x}})\bar{\mu}_i(\tilde{\mathbf{x}})(a_i - r_i)\bigg|\mathbf{x}\right\} + \mathbb{E}\left\{\sum_{i=1}^m f(\tilde{\mathbf{x}})\bar{\mu}_i(\tilde{\mathbf{x}})z_i\bigg|\mathbf{x}\right\} \tag{A.8}$$

Considering the first part in (A.8), we have

$$\mathbb{E}\left\{\sum_{i=1}^m f(\tilde{\mathbf{x}})\bar{\mu}_i(\tilde{\mathbf{x}})(a_i - r_i)\bigg|\mathbf{x}\right\} \tag{A.9}$$

$$\leq f(\mathbf{x})\left(\sum_{i=1}^m \bar{\mu}_i(\mathbf{x})\alpha_i - \sum_{i=1}^m \bar{\mu}_i(\mathbf{x})\mathbb{E}\{r_i|\mathbf{x}\}\right)$$

$$+ \mathbb{E}\left\{\sum_{i=1}^m |f(\tilde{\mathbf{x}})\bar{\mu}_i(\tilde{\mathbf{x}}) - f(\mathbf{x})\bar{\mu}_i(\mathbf{x})||a_i - r_i|\bigg|\mathbf{x}\right\}. \tag{A.10}$$

Since

$$\begin{aligned}
\mathbb{E}\{\mathbf{r}|\mathbf{x}\} &= \boldsymbol{B}\boldsymbol{u}^*(\boldsymbol{x}) \\
&= \boldsymbol{B}\operatorname*{arg\,min}_{\boldsymbol{u}:\mathbb{R}^l_+:\boldsymbol{C}\boldsymbol{u}\leq \mathbf{1}} \langle \boldsymbol{\mu}(\boldsymbol{x}), \boldsymbol{B}\boldsymbol{u}\rangle
\end{aligned}$$

for any stabilizable $\boldsymbol{\alpha}$ we can always find some $\Gamma > 0$, so that

$$\mathbb{E}\left\{\sum_{i=1}^{m} \bar{\mu}_i(\mathbf{x})(\alpha_i - r_i)\bigg|\mathbf{x}\right\} \leq -\Gamma.$$

Hence the first part in (A.10)

$$f(\mathbf{x})\left(\sum_{i=1}^{m}\bar{\mu}_i(\mathbf{x})\alpha_i - \sum_{i=1}^{m}\bar{\mu}_i(\mathbf{x})\mathbb{E}\left\{r_i|\mathbf{x}\right\}\right)$$
$$\leq -\Gamma f(\mathbf{x}).$$

For the second part in (A.10), we define $\Delta\tilde{\mathbf{x}} = \tilde{\mathbf{x}} - \mathbf{x}$. Then

$$f(\mathbf{x} + \Delta\tilde{\mathbf{x}}) - f(\mathbf{x}) = \int_0^1 \bar{\boldsymbol{\mu}}(\mathbf{x} + t\Delta\tilde{\mathbf{x}})\Delta\tilde{\mathbf{x}}dt \leq \int_0^1 \|\Delta\tilde{\mathbf{x}}\|_1 \, dt = \|\Delta\tilde{\mathbf{x}}\|_1$$

Since $a_i$ and $r_i$ are bounded, we choose some $C_3 > 1$ so that $a_i < C_3$ and $r_i < C_3$ for all $i$. Then $\|\Delta\tilde{\mathbf{x}}\|_1$ is bounded by $2mC_3$ and we have

$$|f(\tilde{\mathbf{x}}) - f(\mathbf{x})| < \epsilon_3 f(\mathbf{x})$$

for any given $\epsilon_3 > 0$ and sufficiently large $\|\mathbf{x}\|$. According to Condition 1) in Theorem 4.1, we also have

$$|\bar{\mu}_i(\tilde{\mathbf{x}}) - \bar{\mu}_i(\mathbf{x})| < \epsilon_1.$$

Then if $\|\mathbf{x}\|$ is sufficiently large,

$$\mathbb{E}\left\{\sum_{i=1}^{m}|f(\tilde{\mathbf{x}})\bar{\mu}_i(\tilde{\mathbf{x}}) - f(\mathbf{x})\bar{\mu}_i(\mathbf{x})||a_i - r_i|\bigg|\mathbf{x}\right\}$$
$$\leq 2C_3\mathbb{E}\left\{\sum_{i=1}^{m}(f(\mathbf{x}) + \epsilon_3 f(\mathbf{x}))(\bar{\mu}_i(\mathbf{x}) + \epsilon_1)\bigg|\mathbf{x}\right\}$$
$$- 2C_3\mathbb{E}\left\{\sum_{i=1}^{m}f(\mathbf{x})\bar{\mu}_i(\mathbf{x})\bigg|\mathbf{x}\right\}$$
$$=\underbrace{(2mC_3\epsilon_1 + 2C_3\epsilon_3 + 2mC_3\epsilon_1\epsilon_3)}_{\sigma_1}f(\mathbf{x}) \tag{A.11}$$

holds for any $\epsilon_1, \epsilon_3 > 0$. Hence we have $\sigma_1 \to 0$ when $\|\mathbf{x}\| \to +\infty$.

Now we consider the second part in (A.8).

$$\mathbb{E}\left\{\sum_{i=1}^{m}f(\tilde{\mathbf{x}})\bar{\mu}_i(\tilde{\mathbf{x}})z_i\bigg|\mathbf{x}\right\}$$
$$\leq \mathbb{E}\left\{\sum_{i=1}^{m}f(\mathbf{x})\bar{\mu}_i(\mathbf{x})z_i\bigg|\mathbf{x}\right\} + \mathbb{E}\left\{\sum_{i=1}^{m}|f(\tilde{\mathbf{x}})\bar{\mu}_i(\tilde{\mathbf{x}}) - f(\mathbf{x})\bar{\mu}_i(\mathbf{x})|z_i\bigg|\mathbf{x}\right\} \tag{A.12}$$

For the first part in (A.12), since $z_i \leq r_i$ we have for some $C_4 > 0$ we have

$$\mathbb{E}\left\{z_i(t)\right\} \leq C_4. \tag{A.13}$$

We define the set $\mathcal{G} := \{i : z_i > 0, i \in \mathcal{M}\}$. Since $r_i < C_3$ is bounded by $C_3$, then $x_i < C_3, \forall i \in \mathcal{G}$. If $\|\mathbf{x}\|$ is sufficiently large so that $\|\mathbf{x}\| > mC_3$, we can exclude the case $\mathcal{G} = \mathcal{M}$. According to Condition 2) we have $\bar{\mu}_i(\mathbf{x}) \leq \epsilon_2, \forall i \in \mathcal{G}$ for arbitrarily small $\epsilon_2$. Then

$$\mathbb{E}\left\{\sum_{i \in \mathcal{G}} f(\mathbf{x})\bar{\mu}_i(\mathbf{x})z_i \bigg| \mathbf{x}\right\} < mC_4\epsilon_2 f(\mathbf{x}) \tag{A.14}$$

holds.

Using the same proof method as for (A.11) it can be shown that the second part in (A.12) can be bounded by $\sigma_2 f(\mathbf{x})$ for any $\sigma_2 > 0$.

Define $\theta = \Gamma - \sigma_1 - mC_4\epsilon_2 - \sigma_2$ and choose $\sigma_1, \sigma_2, \epsilon_2$ so that $\theta > 0$ we have the drift

$$\Delta V(\mathbf{x}) \leq -\theta f(\mathbf{x}) \tag{A.15}$$

and which is negative and the Markov chain is f-stable. □

Lemma A.1 is applied to fields which are completely integrable which is of course too restrictive. However, it can be shown that if $\bar{\boldsymbol{\mu}}(\mathbf{x})$ has the properties described in Theorem 4.1, it can be approximated by some (at least piecewise integrable) function $\tilde{\boldsymbol{\mu}}(\mathbf{x})$. The following lemma helps us to achieve our main result.

LEMMA A.2. *If the function $\bar{\boldsymbol{\mu}}(\mathbf{x})$ fulfills the Condition 1), 2) in Theorem 4.1, then there exists a positive, unbounded function $f : \mathbb{R}_+^m \to \mathbb{R}_+$ as given in Definition 3.1, and a positive, continuous, piecewise differentiable function $V : \mathbb{R}_+^m \to \mathbb{R}_+$, such that it holds*

$$\frac{\partial V(\mathbf{x})}{\partial x_i} = f(\mathbf{x})\tilde{\mu}_i(\mathbf{x}), \forall i \in \mathcal{M} \tag{A.16}$$

*on each differentiable subdomain of $V$, and*

$$|\tilde{\mu}_i(\mathbf{x}) - \bar{\mu}_i(\mathbf{x})| < \epsilon_4, \forall i \in \mathcal{M}, \tag{A.17}$$

*for any $\epsilon_4 > 0$ if $\|\mathbf{x}\|$ is sufficiently large.*

*Proof.* In the following we show how to construct the function $V(\mathbf{x})$, $f(\mathbf{x})$ and $\tilde{\boldsymbol{\mu}}(\mathbf{x})$ based on $\bar{\boldsymbol{\mu}}(\mathbf{x})$. Since we only need to ensure that $|\tilde{\mu}_i(\mathbf{x}) - \bar{\mu}_i(\mathbf{x})| < \epsilon_4$ for large $\|\mathbf{x}\|$, it is sufficient to construct the functions on the domain where $\|\mathbf{x}\| \geq B$ for sufficiently large $B$. The function $V$ and $f$ on the domain $\|\mathbf{x}\| \leq B$ can be defined as any positive, bounded, continuously differentiable function, which is continuous on the boundary $\|\mathbf{x}\| = B$.

In the domain of $\|\mathbf{x}\| \geq B$, we at first construct an orthogonal grid such that each cell in the grid is a rectangle (see Fig.A.1 for an example in $m = 3$-dimension). Start by a point $\mathbf{x}^a = \mathbf{X} \in \mathbb{R}_+^m$, the next cell in the dimensions $i, j$ (see Fig.A.2) has the grid points

$$\begin{aligned}
\mathbf{x}^a &= [X_1, ..., X_i, ..., X_j, ..., X_m]^T, \\
\mathbf{x}^b &= [X_1, ..., X_i + \Delta X_i, ..., X_j, ..., X_m]^T, \\
\mathbf{x}^c &= [X_1, ..., X_i, ..., X_j + \Delta X_j, ..., X_m]^T, \\
\mathbf{x}^d &= [X_1, ..., X_i + \Delta X_i, ..., X_j + \Delta X_j, ..., X_m]^T.
\end{aligned}$$

The length of the cell $\Delta X_i, \Delta X_j$ is determined by the equation

$$\begin{aligned}
&\int_0^{\Delta X_i} \bar{\mu}_i(..., x_i, X_j, ...) - \bar{\mu}_i(..., x_i, X_j + \Delta X_j, ...) dx_i \\
&= \int_0^{\Delta X_j} \bar{\mu}_j(..., X_i, x_j, ...) - \bar{\mu}_j(..., X_i + \Delta X_i, x_j, ...) dx_j.
\end{aligned} \tag{A.18}$$

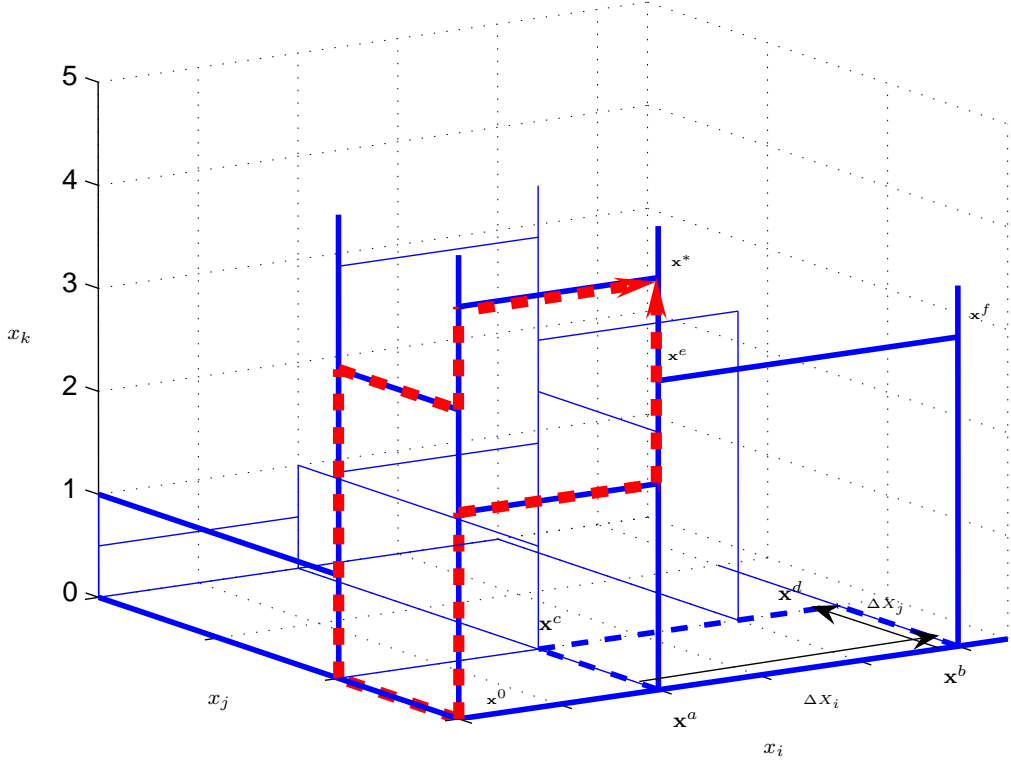

Fig. A.1. *Orthogonal grid (irregular) in $m = 3$-dimension. The line integral between two points on the grid line (e.g. along the two paths marked by dashed line) depends only on the start point $\mathbf{x}^0$ and end point $\mathbf{x}^*$. It is independent of the chosen pathes*

Condition 2) in Theorem 4.1 implies that in the region $\|\mathbf{x}\| \geq B$ for some large constant $B$, the function $\bar{\mu}_i(\mathbf{x})$ decreases with increasing $x_j$ and $\bar{\mu}_j(\mathbf{x})$ decreases with increasing $x_i$ as well. Hence

$$\bar{\mu}_i(..., x_i, X_j, ...) - \bar{\mu}_i(..., x_i, X_j + \Delta X_j, ...) > 0$$
$$\bar{\mu}_j(..., X_i, x_j, ...) - \bar{\mu}_j(..., X_i + \Delta X_i, x_j, ...) > 0$$

and Eqn. (A.18) has positive general solutions with $\Delta X_i, \Delta X_j > 0$. Iteratively take $\boldsymbol{x}^b$, $\boldsymbol{x}^c$ and $\boldsymbol{x}^d$ as start point, we can extend the grid until it covers the subdomain in the dimensions $i, j$. Based on the existing grid lines in the dimensions $i, j$ (e.g. the line $\overline{\mathbf{x}^a \mathbf{x}^b}$ in Fig. A.1), we can repeat the process in a further dimension $k$ and construct the grid in this dimension (the grid $\mathbf{x}^a$-$\mathbf{x}^b$-$\mathbf{x}^e$-$\mathbf{x}^f$). Since relationship of $\Delta X_i$ and $\Delta X_j$ is determined by the definition of $\bar{\mu}_i(\mathbf{x})$ on the particular points, each rectangle in the grid has different height and width so that the constructed grid has a irregular pattern.

Denote the path starting at $\boldsymbol{x}^a$ via $\boldsymbol{x}^b$ to $\boldsymbol{x}^d$ as $S_{abd}$ and the path starting at $\boldsymbol{x}^a$ via $\boldsymbol{x}^c$ to $\boldsymbol{x}^d$ as $S_{acd}$, Eqn. (A.18) ensures that the integral of the function $\bar{\boldsymbol{\mu}}(\mathbf{x})$ along the path $S_{abd}$ equals the integral along the path $S_{acd}$, which is

$$\int_{S_{abd}} \bar{\boldsymbol{\mu}}(\mathbf{x}) \cdot d\mathbf{s}$$
$$= \int_0^{\Delta X_i} \bar{\mu}_i(..., x_i, X_j, ...)dx_i$$
$$+ \int_0^{\Delta X_j} \bar{\mu}_j(..., X_i + \Delta X_i, x_j, ...)dx_j$$
$$= \int_0^{\Delta X_j} \bar{\mu}_j(..., X_i, x_j, ...)dx_j$$
$$+ \int_0^{\Delta X_i} \bar{\mu}_i(..., x_i, X_j + \Delta X_j, ...)dx_i$$
$$= \int_{S_{acd}} \bar{\boldsymbol{\mu}}(\mathbf{x}) \cdot d\mathbf{s}. \tag{A.19}$$

Since Eqn. (A.19) holds for all cells of the grid, the integral between arbitrary two grid points along any grid line has the same value. Hence the vector field $\bar{\boldsymbol{\mu}}(\mathbf{x})$ can be considered as "path-independent" along the grid lines. Then we define a function $f(\mathbf{x})$ whose value on the grid line as the integral of $\bar{\boldsymbol{\mu}}(\mathbf{x})$ along the grid lines, i.e.

$$f(\mathbf{x}^*) := f(\mathbf{X}^0) + \int_S \bar{\boldsymbol{\mu}}(\mathbf{x}) \cdot d\mathbf{s},$$

where $\mathbf{x}^*$ is a point on the grid line and $S$ is an arbitrary path between $\mathbf{x}^*$ and the initial point $\mathbf{X}^0$ along the grid lines.

Define a new vector field by $\boldsymbol{\nu}(\mathbf{x}) := f(\mathbf{x})\bar{\boldsymbol{\mu}}(\mathbf{x})$, the line integral of $\boldsymbol{\nu}(\mathbf{x})$ along the path $S_{abc}$ is

$$\int_{S_{abd}} \boldsymbol{\nu}(\mathbf{x}) \cdot d\mathbf{s} = \int_{S_{abd}} f(\mathbf{x})\bar{\boldsymbol{\mu}}(\mathbf{x}) \cdot d\mathbf{s}$$
$$= \int_{S_{abd}} f(\mathbf{x})df(\mathbf{x})$$
$$= \frac{1}{2}\left(f^2(\mathbf{x}^d) - f^2(\mathbf{x}^a)\right)$$
$$= \int_{S_{acd}} \boldsymbol{\nu}(\mathbf{x}) \cdot d\mathbf{s}.$$

Thus the integral of the vector field $\boldsymbol{\nu}(\mathbf{x})$ between two grid points along the grid lines is also independent of the chosen paths. Then we define a scalar field $V(\mathbf{x})$ whose value on the grid line is given by

$$V(\mathbf{x}^*) := V(\mathbf{X}^0) + \int_S f(\mathbf{x})\bar{\boldsymbol{\mu}}(\mathbf{x}) \cdot d\mathbf{s}.$$

The value of $f(\mathbf{X}^0)$ and $V(\mathbf{X}^0)$ at the initial point $\mathbf{X}^0$ can be chosen as an arbitrary positive constant. Since $\bar{\mu}_i(\mathbf{x}) \geq 0, \forall i \in \mathcal{M}$, we have $f(\mathbf{x}^*) \to +\infty$ and $V(\mathbf{x}^*) \to +\infty$ as $\|\mathbf{x}^*\| \to +\infty$.

Once the value of $V(\mathbf{x}^*)$ is fixed on the grid lines, we obtain the value of $V$ inside a grid cell by the linear interpolation of $V(\mathbf{x}^*)$ along the lines parallel to the diagonal line (see Fig.A.2), i.e. in the lower triangle with $\frac{\Delta x_i}{\Delta X_i} + \frac{\Delta x_j}{\Delta X_j} < 1$, $V$ is defined as

$$V(..., X_i + \Delta x_i, X_j + \Delta x_j, ...) = K_i V(\mathbf{x}^I) + K_j V(\mathbf{x}^J), \tag{A.20}$$

where

$$K_i = \frac{\Delta X_j \Delta x_i}{\Delta X_j \Delta x_i + \Delta X_i \Delta x_j},$$

$$K_j = \frac{\Delta X_i \Delta x_j}{\Delta X_j \Delta x_i + \Delta X_i \Delta x_j},$$

$$\mathbf{x}^I = [X_1, ..., X_i + \Delta x_i + \frac{\Delta X_i}{\Delta X_j}\Delta x_j, X_j, ..., X_m]^T,$$

$$\mathbf{x}^J = [X_1, ..., X_i, X_j + \Delta x_j + \frac{\Delta X_j}{\Delta X_i}\Delta x_i, ..., X_m]^T$$

and in the higher triangle with $\frac{\Delta x_i}{\Delta X_i} + \frac{\Delta x_j}{\Delta X_j} \geq 1$, $V$ is defined as

$$V(..., X_i + \Delta x_i, X_j + \Delta x_j, ...) = K_i V(\mathbf{x}^I) + K_j V(\mathbf{x}^J), \tag{A.21}$$

where

$$K_i = \frac{\Delta X_j \Delta X_i - \Delta X_j \Delta x_i}{2\Delta X_i \Delta X_j - \Delta X_j \Delta x_i - \Delta X_i \Delta x_j},$$

$$K_j = \frac{\Delta X_j \Delta X_i - \Delta X_i \Delta x_j}{2\Delta X_i \Delta X_j - \Delta X_j \Delta x_i - \Delta X_i \Delta x_j},$$

$$\mathbf{x}^I = [..., X_i + \Delta x_i + \frac{\Delta X_i}{\Delta X_j}\Delta x_j - \Delta X_i, X_j + \Delta X_j, ...]^T,$$

$$\mathbf{x}^J = [..., X_i + \Delta X_i, X_j + \Delta x_j + \frac{\Delta X_j}{\Delta X_i}\Delta x_i - \Delta X_j, ...]^T.$$

Eqn. (A.20) and (A.21) determine the value of $V(\mathbf{x})$ on the orthogonal planes stretched by the grid, then the value of $V(\mathbf{x})$ in the space between these planes is calculated by the linear interpolation of the existing value in further dimensions. Similarly, we can also define the value of $f(\mathbf{x})$ in the entire domain.

Observing the function $V(\mathbf{x})$, we can see that it is continuous in $\mathbb{R}_+^m$ and differentiable in each subspace bounded by the grid lines and diagonal lines. For two points $\mathbf{x}$ and $\mathbf{x}'$ which lie in the same cell, under Condition 1) in Theorem 4.1 we have $|\bar{\mu}_i(\mathbf{x}) - \bar{\mu}_i(\mathbf{x}')| \leq \epsilon_1$ and hence $|f(\mathbf{x}) - f(\mathbf{x}')| \leq \epsilon_1 f(\mathbf{x})$ for arbitrarily small $\epsilon_1 > 0$. Then for Eqn. (A.20) it holds

$$V(\mathbf{x}^I) = V(\mathbf{x}^a) + f(\mathbf{x})(\bar{\mu}_i(\mathbf{x}) + \varepsilon_i(\mathbf{x}))\left(\Delta x_i + \frac{\Delta X_i}{\Delta X_j}\Delta x_j\right),$$

$$V(\mathbf{x}^J) = V(\mathbf{x}^a) + f(\mathbf{x})(\bar{\mu}_j(\mathbf{x}) + \varepsilon_j(\mathbf{x}))\left(\Delta x_j + \frac{\Delta X_j}{\Delta X_i}\Delta x_i\right).$$

and further

$$V(\mathbf{x}) = V(\mathbf{x}^a) + f(\mathbf{x})(\bar{\mu}_i(\mathbf{x}) + \varepsilon_i(\mathbf{x}))\Delta x_i + f(\mathbf{x})(\bar{\mu}_j(\mathbf{x}) + \varepsilon_j(\mathbf{x}))\Delta x_j,$$

where the deviation $\varepsilon_i(\mathbf{x}), \varepsilon_j(\mathbf{x}) \to 0$ as $\|\mathbf{x}\| \to +\infty$. Similarly we can also obtain the same result for Eqn. (A.21).

Then the partial derivative of $V$ is

$$\frac{\partial V(\mathbf{x})}{\partial x_i} = f(\mathbf{x})(\bar{\mu}_i(\mathbf{x}) + \epsilon_4).$$

for arbitrarily small $\epsilon_4 > 0$ and we obtain the Lemma A.2. □

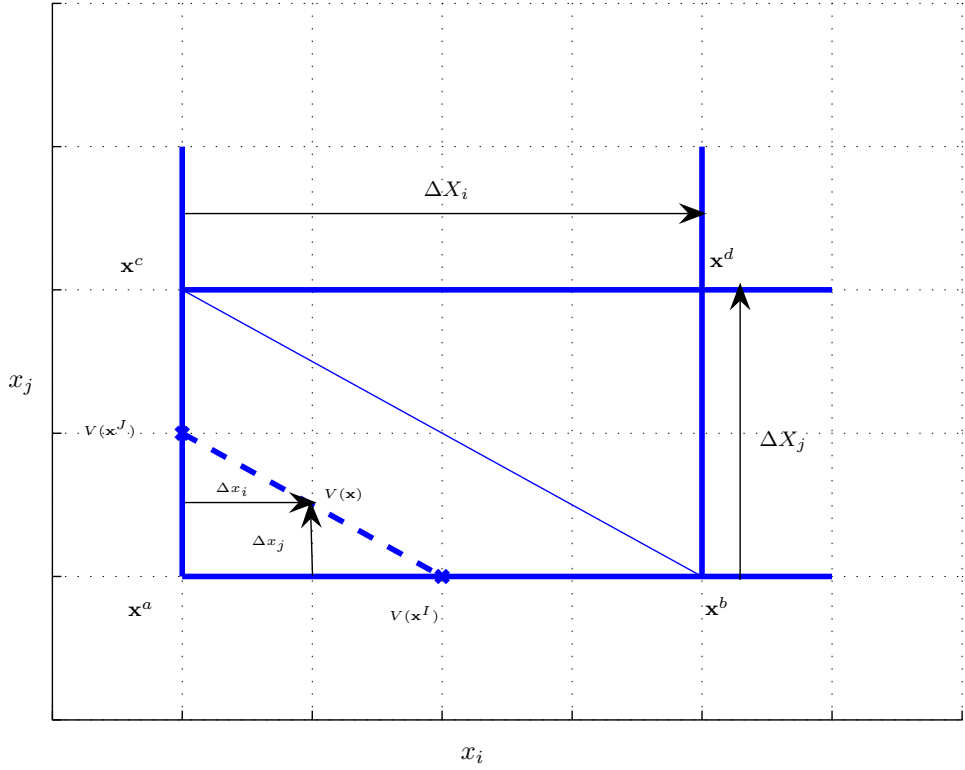

FIG. A.2. *The value of $V(\mathbf{x}) := V(..., X_i + \Delta x_i, X_j + \Delta x_j, ...)$ is calculated by the linear interpolation between the value $V(\mathbf{x}^I)$ and $V(\mathbf{x}^I)$ defined on the grid lines. The line $\overline{\mathbf{x}^I \mathbf{x}^J}$ is parallel to the diagonal $\overline{\mathbf{x}^b \mathbf{x}^c}$*

It can be shown that $f(\mathbf{x})$ and $V(\mathbf{x})$ constructed in Lemma A.2 are positive and grow to infinity as $\|\mathbf{x}\| \to +\infty$. Now we use the function $V(\mathbf{x})$ and $f(\mathbf{x})$ in Lemma A.2 as the Lyapunov function and the stability measure, respectively. It can also be shown that $\Delta V(\mathbf{x})$ is bounded if $\mathbf{x}$ lies in some compacted region $\mathcal{B}$ and the arrival rates $a_i$ and transmission rates $r_i$ are bounded. Hence the Lyapunov condition (A.1) is satisfied.

Next we consider the drift $\Delta V(\mathbf{x})$ in Lyapunov condition (A.2) where $\mathbf{x} \in \mathcal{B}^c$. The connection between $\mathbf{x}$ and $\mathbf{x} + \Delta \boldsymbol{x}$ probably pass through multiple differentiable subspaces of $V(\mathbf{x})$ (see Fig.A.4), so we denote the intersection of the connecting line and the boundary of the subspaces as $\mathbf{x}^{(1)}, ..., \mathbf{x}^{(L)}$ and the difference as $\Delta \mathbf{x}^{(1)} = \mathbf{x}^{(1)} - \mathbf{x}, ..., \Delta \mathbf{x}^{(l)} = \mathbf{x}^{(l+1)} - \mathbf{x}^{(l)}$. The drift is written as

$$\Delta V(\mathbf{x}) = \mathbb{E}\left\{ V(\mathbf{x} + \Delta \boldsymbol{x}) - V(\mathbf{x}^{(L)}) + \sum_{l=2}^{L} V(\mathbf{x}^{l+1}) - V(\mathbf{x}^{(l)}) + V(\mathbf{x}^{(1)}) - V(\mathbf{x}) \bigg| \mathbf{x} \right\}$$

$$= \mathbb{E}\left\{ \sum_{l=1}^{L+1} f(\tilde{\mathbf{x}}^{(l)}) \tilde{\boldsymbol{\mu}}(\tilde{\mathbf{x}}^{(l)}) \cdot \Delta \mathbf{x}^{(l)} \bigg| \mathbf{x} \right\}$$

$$\leq \mathbb{E}\left\{ \sum_{l=1}^{L+1} f(\tilde{\mathbf{x}}^{(l)}) \bar{\boldsymbol{\mu}}(\tilde{\mathbf{x}}^{(l)}) \cdot \Delta \mathbf{x}^{(l)} + \epsilon_4 \left\| \Delta \mathbf{x}^{(l)} \right\| f(\tilde{\mathbf{x}}^{(l)}) \bigg| \mathbf{x} \right\},$$

where $\tilde{\mathbf{x}}^{(l)}$ is some point in the $l$-th subspace. Since the arrival rates $a_i$ and the transmission rates

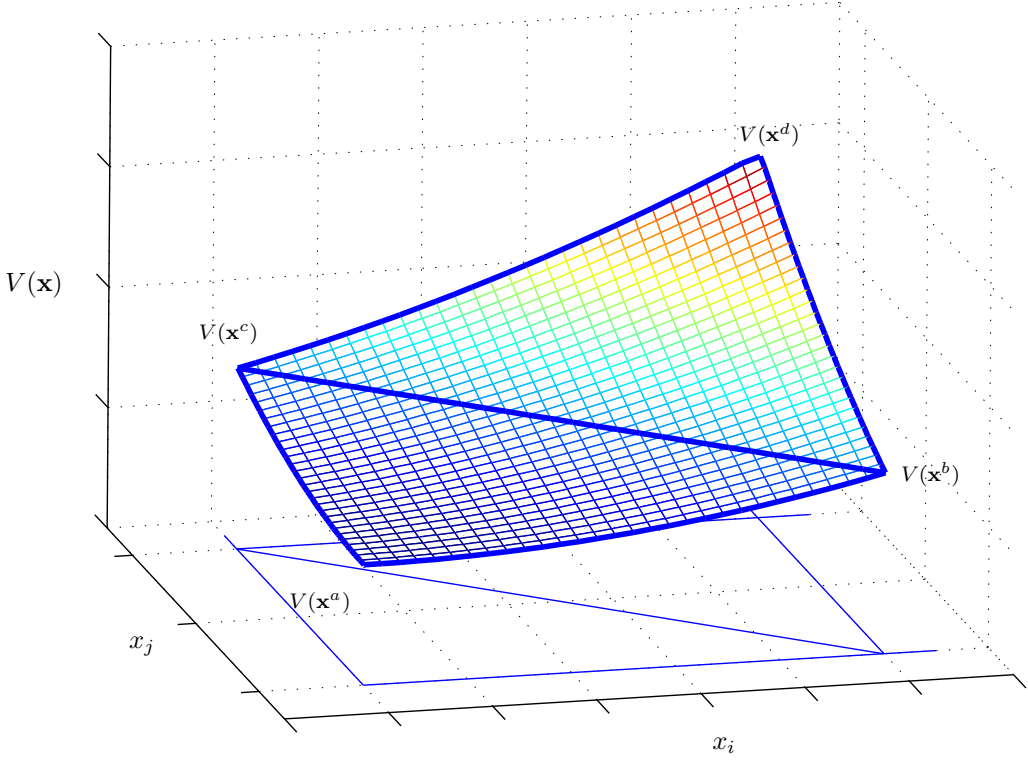

Fig. A.3. *The Lyapunov function $V(\mathbf{x})$ is differentiable inside the subdomain between $\mathbf{x}^a$, $\mathbf{x}^b$, $\mathbf{x}^c$ and the subdomain between $\mathbf{x}^b$, $\mathbf{x}^c$, $\mathbf{x}^d$*

$r_i$ are bounded for all $i \in \mathcal{M}$, the difference $\|\Delta \mathbf{x}\|$ is bounded. Thus according to Condition 1) in Theorem 4.1 we have $\left|\bar{\mu}_i(\tilde{\mathbf{x}}^{(l)}) - \bar{\mu}_i(\tilde{\mathbf{x}}^{(1)})\right| < \epsilon_1$ and $\left|f(\tilde{\mathbf{x}}^{(l)}) - f(\tilde{\mathbf{x}}^{(1)})\right| < \epsilon_1 f\left(\tilde{\mathbf{x}}^{(1)}\right)$ for arbitrary $\epsilon_1 > 0$ if $\|\tilde{\mathbf{x}}^{(1)}\|$ is large. The drift

$$\Delta V(\mathbf{x})$$
$$\leq \mathbb{E}\left\{ f(\tilde{\mathbf{x}}^{(1)})\bar{\boldsymbol{\mu}}(\tilde{\mathbf{x}}^{(1)}) \cdot \sum_{l=1}^{L+1} \Delta \mathbf{x}^{(l)} + \sigma_3 f(\mathbf{x}) \middle| \mathbf{x} \right\}$$
$$\leq \mathbb{E}\left\{ f(\tilde{\mathbf{x}}^{(1)})\bar{\boldsymbol{\mu}}(\tilde{\mathbf{x}}^{(1)}) \cdot (\mathbf{x} + \Delta \boldsymbol{x} - \mathbf{x}) \middle| \mathbf{x} \right\} + \sigma_3 f(\mathbf{x}),$$

where $\sigma_3$ is some small constant.

Using the previous result in (A.15), it holds

$$\Delta V(\mathbf{x})$$
$$\leq \mathbb{E}\left\{ \sum_{i=1}^{m} f(\tilde{\mathbf{x}})\bar{\mu}_i(\tilde{\mathbf{x}})\left(a_i - r_i + z_i\right) \middle| \mathbf{x} \right\} + \sigma_3 f(\mathbf{x})$$
$$\leq -\theta f(\mathbf{x}) + \sigma_3 f(\mathbf{x})$$
$$\leq -\theta' f(\mathbf{x})$$

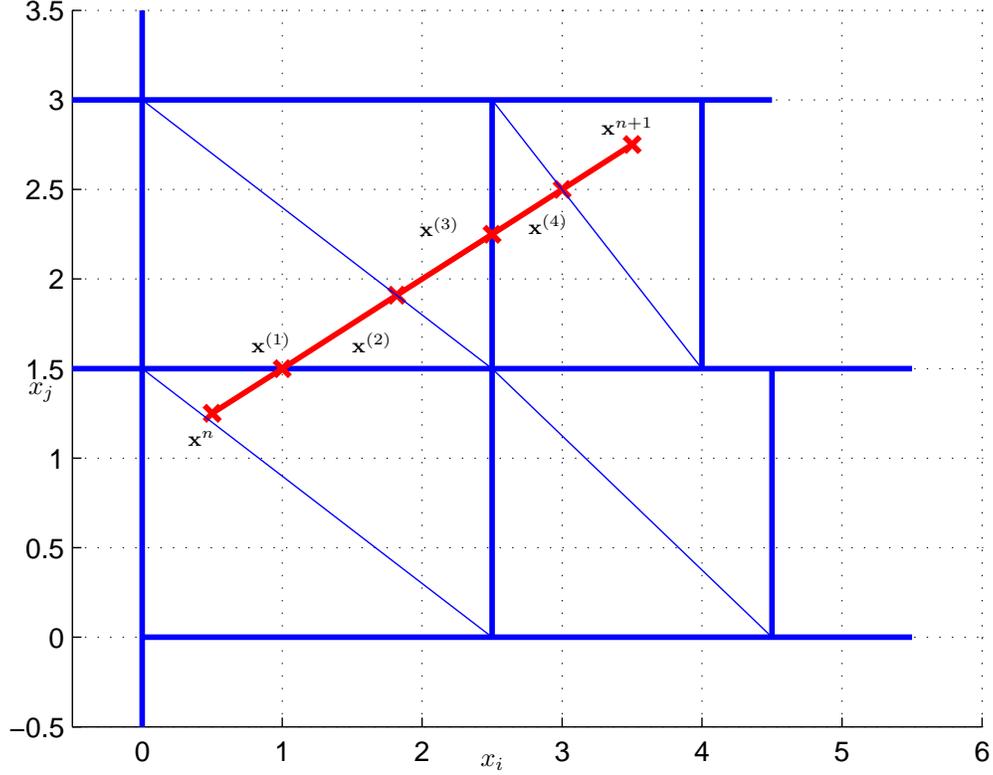

Fig. A.4. *The drift $\Delta V$ crosses 5 subdomains, which can be written as the sum of the difference between $V(\mathbf{x})$, $V(\mathbf{x}^{(1)})$, ..., and $V(\mathbf{x} + \Delta \boldsymbol{x})$*

for some $\theta' > 0$ if $\|\mathbf{x}\| > B$, for some $B > 0$. The drift is negative thus the Markov chain is positive recurrent.

At last, we prove that the chain is also f-stable for the magnitude function $f(\mathbf{x})$. We can write

$$\begin{aligned}
&\mathbb{E}\left\{V(\mathbf{x} + \Delta \boldsymbol{x}) | \mathbf{x}\right\} \\
\leq &\mathbb{E}\left\{V(\mathbf{x} + \Delta \boldsymbol{x}) | \mathbf{x} > B\right\} \Pr(\mathbf{x} > B) + \mathbb{E}\left\{V(\mathbf{x} + \Delta \boldsymbol{x}) | \mathbf{x} \leq B\right\} \Pr(\mathbf{x} \leq B) \\
\leq &\mathbb{E}\left\{V(\mathbf{x}) - \theta' f(\mathbf{x}) | \mathbf{x} > B\right\} \Pr(\mathbf{x} > B) + \mathbb{E}\left\{V(\mathbf{x} + \Delta \boldsymbol{x}) | \mathbf{x} \leq B\right\} \Pr(\mathbf{x} \leq B) \\
\leq &\mathbb{E}\left\{V(\mathbf{x})\right\} - \theta' f(\mathbf{x}) + C_5,
\end{aligned}$$

where $C_5$ is some constant satisfying

$$\begin{aligned}
C_5 \geq &\mathbb{E}\left\{V(\mathbf{x} + \Delta \boldsymbol{x}) | \mathbf{x} \leq B\right\} \Pr(\mathbf{x} \leq B) \\
&+ \mathbb{E}\left\{\theta' f(\mathbf{x}) | \mathbf{x} \leq B\right\} \Pr(\mathbf{x} \leq B).
\end{aligned}$$

Using the telescoping machinery, the summation of the drift over $T$ time slots yields

$$\mathbb{E}\left\{V(\mathbf{x}^T)\right\} \leq \mathbb{E}\left\{V(\mathbf{x}^1)\right\} - \theta' \sum_{n=1}^{T} \mathbb{E}\left\{f(\mathbf{x})\right\} + T \cdot C_5,$$

and since $V(\mathbf{x})$ is non-negative function, it holds

$$\sum_{n=1}^{T} \mathbb{E}\left\{f(\mathbf{x})\right\} \leq \frac{\mathbb{E}\left\{V(\mathbf{x}^1)\right\}}{\theta'} + \frac{T \cdot C_5}{\theta'}.$$

Hence we have

$$\limsup_{t \to +\infty} \frac{1}{T} \sum_{n=1}^{T} \mathbb{E}\left\{f(\mathbf{x})\right\} \leq \frac{\mathbb{E}\left\{V(\mathbf{x}^1)\right\}}{T\theta'} + \frac{C_5}{\theta'} < +\infty$$

which completes the proof.

**Appendix B. Proof of Corollary 4.2.**

By Condition 2) of Corollary 4.2 we can assume that the random walk evolves on $\mathbb{R}_+^m$. Hence, we can skip Condition 2) of Theorem 4.1 since this condition (as its counterpart in Corollary 4.2) ensures positivity of the random walk. We need to show that from

$$\|\nabla \log \mu_i(\boldsymbol{x})\| \leq \epsilon, \quad \forall i \in \mathcal{M}, \ \|\boldsymbol{x}\| > C_6(\epsilon), \tag{B.1}$$

(where $C_6(\epsilon)$ is sufficiently large) it follows:

$$\left| \frac{\mu_i(\boldsymbol{x} + \Delta\boldsymbol{x})}{\sum_{j \in \mathcal{M}} \mu_j(\boldsymbol{x} + \Delta\boldsymbol{x})} - \frac{\mu_i(\boldsymbol{x})}{\sum_{j \in \mathcal{M}} \mu_j(\boldsymbol{x})} \right| \leq \epsilon \tag{B.2}$$

For orientation, let us assume more restrictive conditions first: take $\mu_i$, $\forall i \in \mathcal{M}$ Lipschitz continuous and let $\sum_{j \in \mathcal{M}} \mu_j(\boldsymbol{x}) \to \infty$ if $\|\boldsymbol{x}\| \to \infty$. Note, that these conditions already encompasses Meyn's perturbation (3.7) together with e.g. a linear cost function.

It is easy to prove the corollary with these assumptions: by the mean value theorem we have

$$\mu_i(\boldsymbol{x} + \Delta\boldsymbol{x}) = \mu_i(\overline{\boldsymbol{x}}) + \nabla_{\boldsymbol{x}}^T \mu_i(\widetilde{\boldsymbol{x}}) \Delta\overline{\boldsymbol{x}}$$

where $\overline{\boldsymbol{x}}$ is an (arbitrary) point on line connecting $\boldsymbol{x}$ and $\boldsymbol{x} + \Delta\boldsymbol{x}$ whereas $\widetilde{\boldsymbol{x}}$ is a point connecting $\overline{\boldsymbol{x}} + \Delta\overline{\boldsymbol{x}}$. Since the field is Lipschitz we have $\nabla_{\boldsymbol{x}}^T \mu_i(\overline{\boldsymbol{x}}) \leq C_7$ uniformly. Furthermore, since the policy is non-idling $\sum_{j \in \mathcal{M}} \mu_j(\boldsymbol{x} + \Delta\boldsymbol{x}) \geq C_8$ where the normalization constant $C_8$ can be chosen as large as possible without altering the policy (by the construction of the policy). Moreover, since $\sum_{j \in \mathcal{M}} \mu_j(\boldsymbol{x}) \to \infty, \|\boldsymbol{x}\| \to \infty$, condition (B.2) is equivalent to

$$|\mu_i(\boldsymbol{x} + \Delta\boldsymbol{x}) - \mu_i(\boldsymbol{x})| \leq \epsilon \sum_{j \in \mathcal{M}} \mu_j(\overline{\boldsymbol{x}})$$

and, again, by the mean value theorem:

$$\left| \nabla^T \mu_i(\overline{\boldsymbol{x}}) \Delta\boldsymbol{x} \right| \leq \epsilon \sum_{j \in \mathcal{M}} \mu_j(\overline{\boldsymbol{x}})$$

Here, we tacitly assumed that we have selected $\overline{\boldsymbol{x}}$ accordingly. Since $\Delta\boldsymbol{x}$ is fixed and by the positivity of $\mu_i$ it is sufficient that

$$\|\nabla \mu_i(\boldsymbol{x})\| \leq \frac{\epsilon}{\|\Delta\boldsymbol{x}\|} \mu_i(\boldsymbol{x})$$

which is equivalent to condition (B.1) with some $\|\boldsymbol{x}\| > C_6(\epsilon')$ ($\epsilon'$ slightly smaller).

Let us now prove the general case. Condition (B.1) can be written as

$$\frac{1}{\sum_{j \in \mathcal{M}} \mu_j(\boldsymbol{x})} \nabla^T \mu_i(\boldsymbol{x}) \Delta\boldsymbol{x} = \epsilon_n,$$

for some $\bm{x}$ with $\|\bm{x}\| > C(\epsilon_n)$ where $\epsilon_n$ is a zero sequence and $C(\epsilon_n)$ is strictly increasing for any fixed $\Delta \bm{x} \in \mathbb{R}^m$. Now, again, by the mean value theorem

$$\left| \frac{\mu_i(\bm{x} + \Delta \bm{x})}{\sum_{j \in \mathcal{M}} \mu_j(\bar{\bm{x}}) + \nabla^T \mu_j(\tilde{\bm{x}}) \Delta \bm{x}} - \frac{\mu_i(\bm{x})}{\sum_{j \in \mathcal{M}} \mu_j(\bar{\bm{x}}) - \nabla^T \mu_j(\tilde{\underline{\bm{x}}}) \Delta \underline{\bm{x}}} \right| \leq \epsilon, \tag{B.3}$$

where we set $\bar{\bm{x}}$ as before and let $\bm{x} + \Delta \underline{\bm{x}} = \bar{\bm{x}}$ and $\bar{\bm{x}} + \Delta \bar{\bm{x}} = \bm{x} + \Delta \bm{x}$. $\tilde{\underline{\bm{x}}}, \tilde{\bm{x}}$ are points on the line connecting $\bm{x}$ and $\bar{\bm{x}}$ respectively $\bar{\bm{x}}$ and $\bm{x} + \Delta \bm{x}$. Note that $\mu_j(\bar{\bm{x}})$ is zero if and only if $\mu_i(\bm{x} + \Delta \bm{x})$ and $\mu_i(\bm{x})$ are both zero since otherwise by condition (B.1) the gradient would be zero as well. Since in this case the condition is trivially satisfied so that we exclude it.

Hence from (B.3) it follows

$$\left| \mu_i(\bm{x} + \Delta \bm{x}) - \mu_i(\bm{x}) \frac{\sum_{j \in \mathcal{M}} \mu_j(\bar{\bm{x}})(1 + \overbrace{\frac{\nabla^T \mu_j(\tilde{\bm{x}}) \Delta \bar{\bm{x}}}{\mu_j(\bar{\bm{x}})}}^{(A)})}{\sum_{j \in \mathcal{M}} \mu_j(\bar{\bm{x}})(1 - \underbrace{\frac{\nabla^T \mu_j(\tilde{\underline{\bm{x}}}) \Delta \underline{\bm{x}}}{\mu_j(\bar{\bm{x}})}}_{(B)})} \right| \leq \epsilon \cdot \sum_{j \in \mathcal{M}} \mu_j(\bar{\bm{x}}) \left( 1 + \frac{\nabla^T \mu_j(\tilde{\bm{x}}) \Delta \bar{\bm{x}}}{\mu_j(\bar{\bm{x}})} \right).$$

We can prove that, because of condition (B.1), (A) and (B) are zero sequences: suppose $\nabla^T \mu_j(\tilde{\bm{x}})$ is non-zero (then we can stop anyway) then by the repeated application of the mean value theorem, denominator of, say, (A) can be written as:

$$\mu_j(\bar{\bm{x}}) = \mu_j(\tilde{\bm{x}}) + \nabla \mu_j(\bm{x}_2) \Delta \bm{x}_2$$

This process generates sequences in $\mathbb{R}_+^m$ with $\tilde{\bm{x}} = \bm{x}_1, \bm{x}_2, ...$ and $\Delta \bar{\bm{x}} = \Delta \bar{\bm{x}}_1 \subset \Delta \bar{\bm{x}}_2, ...$ which are bounded and hence we can pick subsequences converging to some set of limit points $\bm{x}_\infty^{(k)}, k = 1, 2, ...$. Note that we can restrict the number of limit points to at most two since by defintion every limit point is visited arbitrarily often and infinitely close and by construction of the sequence there is no possibility of more than two limit points which neither contain the other in between them. Take these two limit points with corresponding subsequence $\bm{x}_n^{(k)}, k = 1, 2$: by continuous differentiability we have $\mu_j(\bm{x}_n^{(k)}) \to \mu_j(\bm{x}_\infty^{(k)})$ and $\nabla \mu_j(\bm{x}_n^{(k)}) \to \nabla \mu_j(\bm{x}_\infty^{(k)}), k = 1, 2$. It must also hold in the limit:

$$\mu_j(\bm{x}_\infty^{(1)}) + \nabla^T \mu_j(\bm{x}_\infty^{(2)})(\bm{x}_\infty^{(1)} - \bm{x}_\infty^{(2)}) = \mu_j(\bm{x}_\infty^{(2)})$$

(and vice versa). Since then

$$\frac{\nabla^T \mu_j(\bm{x}_\infty^{(2)})(\bm{x}_\infty^{(1)} - \bm{x}_\infty^{(2)})}{\mu_j(\bm{x}_\infty^{(2)}) \mu_j(\bm{x}_\infty^{(2)})} \leq \epsilon,$$

(and vice versa) where $\epsilon > 0$ is arbitrarily small by condition (B.1) we conclude that $\mu_j(\bm{x}_\infty^{(1)}) = \mu_j(\bm{x}_\infty^{(2)})$ (but not necessarily $\bm{x}_\infty^{(1)} = \bm{x}_\infty^{(2)}$).

Now, we can proceed the process sufficiently often as

$$\frac{\nabla^T \mu_j(\bm{x}_1) \Delta \bm{x}_1}{\mu_j(\bar{\bm{x}})}$$
$$\leq \frac{\nabla^T \mu_j(\bm{x}_1) \Delta \bm{x}_1}{\mu_j(\bm{x}_1) \left( 1 + \frac{\nabla^T \mu_j(\bm{x}_2) \Delta \bm{x}_2}{\mu_j(\bm{x}_1)} \right)}$$
$$\leq ...$$

such that in the final step

$$\frac{\nabla^T \mu_j(\boldsymbol{x}_{n+1})\Delta \boldsymbol{x}_{n+1}}{\mu_j(\boldsymbol{x}_n)} = \frac{(\nabla^T \mu_j(\boldsymbol{x}_\infty^{(k)}) + \epsilon_n^k)\Delta \boldsymbol{x}_{n+1}}{\mu_j(\boldsymbol{x}_\infty^{(l)}) + \epsilon_n^l}$$
$$= \frac{(\nabla^T \mu_j(\boldsymbol{x}_\infty^{(k)}) + \epsilon_n^k)\Delta \boldsymbol{x}_{n+1}}{\mu_j(\boldsymbol{x}_\infty^{(k)}) + \epsilon_n^l}$$
$$\leq \epsilon, \ k, l = 1, 2,$$

by condition (B.1). Hence, we have

$$\frac{\sum_{j \in \mathcal{M}} \mu_j(\bar{\boldsymbol{x}}) \left(1 + \frac{\nabla^T \mu_j(\bar{\boldsymbol{x}})\Delta \bar{\boldsymbol{x}}}{\mu_j(\bar{\boldsymbol{x}})}\right)}{\sum_{j \in \mathcal{M}} \mu_j(\bar{\boldsymbol{x}}) \left(1 + \frac{\nabla^T \mu_j(\bar{\boldsymbol{x}})\Delta \boldsymbol{x}}{\mu_j(\bar{\boldsymbol{x}})}\right)} = \frac{(1 + \epsilon_n')}{(1 + \epsilon_n'')}$$
$$= 1 + \epsilon_n''', \quad \epsilon_n', \epsilon_n'' \text{ zero sequences}$$

and further

$$|\mu_i(\boldsymbol{x} + \Delta \boldsymbol{x}) - \mu_i(\boldsymbol{x})(1 + \epsilon_n''')| \leq |\mu_i(\boldsymbol{x} + \Delta \boldsymbol{x}) - \mu_i(\boldsymbol{x})| + \mu_i(\boldsymbol{x})\epsilon_n'''$$
$$\leq \epsilon \sum_{j \in \mathcal{M}} \mu_j(\tilde{\boldsymbol{x}})(1 + \epsilon_n'),$$

which is equivalent to:

$$|\mu_i(\boldsymbol{x} + \Delta \boldsymbol{x}) - \mu_i(\boldsymbol{x})| \leq \epsilon \sum_{j \in \mathcal{M}} \mu_j(\bar{\boldsymbol{x}})(1 + \epsilon_n') - \epsilon_n''\mu_i(\boldsymbol{x}).$$

Since $\bar{\boldsymbol{x}}$ is arbitrary and can be suitably choosen, condition (B.1) with some $\|\boldsymbol{x}\| > C_6 \ (\epsilon'''')$ is sufficient for the latter to hold.

**Appendix C. Proof of Corollary 4.3.**

We can write

$$\mu_i(\boldsymbol{x}) = \frac{\partial h}{\partial x_i}(\boldsymbol{x}) = l(x_i)\frac{\partial h_0}{\partial \tilde{x}_i}(\tilde{\boldsymbol{x}})$$

where we defined $l := \frac{\partial \tilde{x}_i}{\partial x_i}$. Note, that here $\tilde{x}_i$ only depends on $x_i$. The gradient of the weight $\mu_i(\boldsymbol{x})$ is given by:

$$\frac{\partial \mu_i}{\partial x_j}(\boldsymbol{x}) = \begin{cases} \frac{\partial l}{\partial x_i}(x_i)\frac{\partial h_0}{\partial \tilde{x}_i}(\tilde{\boldsymbol{x}}) + l(x_i)\frac{\partial}{\partial x_i}\frac{\partial h_0}{\partial \tilde{x}_i}(\tilde{\boldsymbol{x}}) & i = j \\ \frac{\partial}{\partial x_j}\frac{\partial h_0}{\partial \tilde{x}_i}(\tilde{\boldsymbol{x}}) \cdot l(x_i) & i \neq j \end{cases}$$

Define $\boldsymbol{x}^\Delta := \boldsymbol{x} + \Delta\boldsymbol{x}$ and $\tilde{\boldsymbol{x}}^\Delta := \tilde{\boldsymbol{x}}(\boldsymbol{x}^\Delta)$. From the proof of Corollary 4.2 it is clear that we only have to show that

$$\frac{\left|\nabla^T \mu_i(\boldsymbol{x})\Delta \boldsymbol{x}\right|}{\|\boldsymbol{\mu}(\boldsymbol{x}^\Delta)\|} \leq \epsilon,$$

for some $\epsilon > 0$ arbitrarily small. This can be rewritten as:

$$\frac{\frac{\partial l}{\partial x_i}(x_i)\frac{\partial h_0}{\partial \tilde{x}_i}(\tilde{\boldsymbol{x}})\Delta x_i + l(x_i)\frac{\partial}{\partial x_i}\frac{\partial h_0}{\partial \tilde{x}_i}(\tilde{\boldsymbol{x}})\Delta x_i}{\sum_{j \in \mathcal{M}} l(x_j^\Delta)\frac{\partial h_0}{\partial \tilde{x}_j}(\tilde{\boldsymbol{x}}^\Delta)} + \frac{l(x_i)\sum_{j \in \mathcal{M}, j \neq i}\frac{\partial}{\partial x_j}\frac{\partial h_0}{\partial \tilde{x}_i}(\tilde{\boldsymbol{x}})\Delta x_j}{\sum_{j \in \mathcal{M}} l(x_j^\Delta)\frac{\partial h_0}{\partial \tilde{x}_j}(\tilde{\boldsymbol{x}}^\Delta)} \leq \epsilon$$

Since $\frac{\partial h_0}{\partial \tilde{x}_i}, l$ are Lipschitz, thus $\frac{\partial}{\partial x_j}\frac{\partial h_0}{\partial \tilde{x}_i}, \frac{\partial l}{\partial x_i}$ are uniformly bounded, and $l(x_i), \frac{\partial h_0}{\partial \tilde{x}_i}(\tilde{\boldsymbol{x}}) \geq l^{1+\epsilon}(x_i) \to \infty$ when $x_i \to \infty$, the effect of $\Delta \boldsymbol{x}$ vanishes in the denominator. The condition $\frac{\partial h_0}{\partial \tilde{x}_i}(\tilde{\boldsymbol{x}}) \geq l^{1+\epsilon}(x_i)$ is required since we have expressions of the form

$$\frac{l(x_i)l(x_j)}{l(x_i)\frac{\partial h_0}{\partial \tilde{x}_i}(\tilde{\boldsymbol{x}}) + l(x_j)\frac{\partial h_0}{\partial \tilde{x}_j}(\tilde{\boldsymbol{x}})}$$

which then become arbitrarily small.